\documentclass[twocolumn,preprintnumbers,amsmath,amssymb]{revtex4}

\usepackage{graphicx}
\usepackage{dcolumn}
\usepackage{bm}
\newcommand \be  {\begin{equation}}
\newcommand \bea {\begin{eqnarray} \nonumber }
\newcommand \ee  {\end{equation}}
\newcommand \eea {\end{eqnarray}}

\begin{document}


\title{Glassy dynamics: mean-field landscape pictures\\ versus growing length scale scenarii}

\author{Jean-Philippe Bouchaud}
\email{bouchau@spec.saclay.cea.fr}
 \affiliation{$^{*}$Service de Physique de l'{\'E}tat Condens{\'e}, Orme
 des Merisiers, 
 CEA Saclay, 91191 Gif sur Yvette Cedex, France;\\
 and Science \& Finance, Capital Fund Management, 6-8 Bd Haussmann,
 75009 Paris, France.\\
 }%

 \date{\today}
\begin{abstract}
This is a short review on the compatibility between (a) mean-field, mode-coupling theories of the glass transition, 
where potential energy landscape ideas are natural, and (b) the necessity of describing the slowing down of glassy
materials in terms of a growing cooperative length, absent from mean-field descriptions. We summarize 
some of the outstanding questions that remain before we can say we understand why glasses do not flow.
 \end{abstract}
 
 \maketitle
 
\section{The problem}

The most salient properties of {\it fragile} glasses are (a) the non exponential (``stretched")
and heterogeneous nature of the relaxation; (b) the extremely fast rise of their viscosity $\eta$ 
that increases by 15 orders of magnitude as the temperature is decreased
by less than a factor 2, and appears to diverge at a finite (Vogel-Fulcher) temperature; (c) 
the aging and memory effects of the out-of-equilibrium phase, that shows some similarities with
spin-glasses. A rather remarkable aspect of the Vogel-Fulcher divergence (b) is that
the extrapolated freezing temperature $T_{VF}$ is found to be very close, for a whole range of materials, 
to the Kauzmann temperature $T_K$ where the extrapolated entropy of the glass becomes
smaller than that of the crystal. In other words, the viscosity of 
glasses (a dynamical property) seems highly correlated to the number of microscopic 
configurations in which the glass can get stuck. The smaller this number, the larger the viscosity. 
The striking observation that makes the `problem of glasses' interesting is that very many, totally different
materials, exhibit the same properties, pointing to the existence of a somewhat universal mechanism: glassy
dynamics is physics more than chemistry. 

Two apparently quite different frameworks have been discussed in the (fourty years old) literature to
account for this phenomenology: 1) phase space/energy landscape pictures, where the system is trapped 
in metastable states of varying depth. The dynamics is made of small harmonic vibrations around each
metastable configurations, separated by hops between different 
minima of the free-energy; 2) cooperatively rearranging regions of
increasing length. The dynamics becomes sluggish because larger and larger regions of the
material have to move simultaneously to allow substantial motion of individual particles. 
Although this idea seems most reasonable, its reality has remained elusive until recently: a consistent 
definition of this growing length, its experimental measurement and its calculation within a theoretical model
(even highly simplified) are subjects of topical activity. Interestingly, similar concepts are also relevant
for the description of other ``jammed'' systems, such as dense granular assemblies that flow in a very jerky way.

The landscape picture can be given more flesh within mean-field theories. The Random Energy Model, for 
example, contains already a lot of the glass phenomenology (entropy crisis at $T_K$, aging..). More elaborate 
mean field spin-glasses theories lead to dynamical equations that are identical to the 
``Mode Coupling" theory (MCT) of supercooled liquids. MCT is considered by many to be the only available first principle theory 
of the supercooled state, starting from interacting atoms and making its way up to
compute the viscosity of the liquid as a function of density and temperature. This theory makes a number of 
quantitative predictions that can be compared to experiments, some of which in remarkable agreement with observations. 
This analogy between MCT and mean-field spin-glasses allows one to interpret the MCT scenario for 
dynamical arrest in a clear fashion: the potential energy landscape has only 
unstable saddle points above a certain ``threshold" energy, around which the 
system can only pause momentarily before continuing its exploration of
phase space. This corresponds to the high temperature liquid phase. 
At the energy threshold (corresponding to $T_{MCT}$), the saddles only
have marginal (zero curvature) escape directions, 
responsible for the MCT divergence of relaxation 
times. For lower energies (temperature) there are only minima where the 
system gets trapped. The number of these minima is exponential 
in the size of the system as long as $T_K < T < T_{MCT}$, but the
barriers between them are (in mean-field) infinite: once trapped in a minimum, the system remains there forever.

Is this mean-field picture at all relevant for finite range interactions?  Phase-space pictures cannot be directly 
applied when the dynamics is local: the energy surface must somehow ``factorize''. There cannot be a diverging time scale 
without a diverging {\it length} scale. ``Hops'' in phase space should correspond to definite spatial structures 
(vacancies? strings? fractal clusters?). The observed dynamical heterogeneities, and the corresponding 
viscosity/diffusion decoupling must be accounted for. 
Hidden behind phase space pictures, there must thus be a 
dynamical length scale $\xi(T)$ governing the slowing down of these materials. Contrarily to
simple systems where this length scale is the characteristic size over which some order (ferromagnetic, crystalline, etc.)
is established, the difficulty of glasses and spin-glasses is that no obvious local order sets in. The 
definition of a dynamical length scale is more subtle and requires a {\it four point} density correlation function 
that attempts to quantify the cooperativity of the dynamics. This four-point function plays the role of
the usual two point correlation in second order phase transitions. Following this path, one can
show that MCT in fact predicts the divergence of $\xi(T)$ when $T \to T_{MCT}$; correspondingly, fluctuations 
become dominant below six dimensions, so that MCT cannot be quantitative in three dimensions. The second, perhaps 
more essential, problem of MCT is that barriers between metastable states must be finite for realistic potentials, 
even for $T < T_{MCT}$. Therefore, the predicted divergence of the viscosity at $T_{MCT}$ is in general smeared out.

What is then the nature of the growing length scale at lower temperatures? Old free volume ideas, recently revived
within the context of `facilitated' models, suggest that mobility defects trigger the dynamics and become more and
more dilute at temperature decreases -- the length scale is then the distance between defects. A more ambitious
scenario, proposed by Adam-Gibbs in the 60's and, inspired by mean-field spin-glasses, by Kirkpatrick, Thirumalai 
and Wolynes in the 80's, relates the size of collectively rearranging regions to the configurational entropy of 
the glass. The idea is that of entropic melting of frozen clusters: small clusters have few low energy configurations and
are pinned in one of them by the external environment; large clusters can explore many configurations and free
themselves from any boundary conditions. At small scales, the dynamics from state to state is fast (low barriers) but 
leads to nowhere -- the system ends up always visiting the same state. For larger scales, the system can at last 
delocalize itself in phase space and kill correlations, but this takes an 
increasingly large time. The crossover scale $\xi$ then sets the relaxation 
time and diverges when the entropy goes to zero, explaining the deep 
connection between dynamics and thermodynamics (absent in mobility defect theories). The supercooled liquid is in a
{\it mosaic state}, made up as a patchwork of all possible frozen configurations, with cell size $\xi$.
The dynamics within scale $\xi$ is collective and landscape (trap) pictures
should be relevant. For larger scales, however, the dynamics occurs in parallel and global phase space ideas are meaningless.
The mosaic state offers a plausible interpretation of the random first order transition to an amorphous 
solid predicted by mean-field theories.  The relevance of this beautiful scenario, where the liquid slows down because 
of the emergence of a very large number of metastable
states that momentarily trap the system, with larger and larger frozen regions as the system is allowed to 
visit deeper and deeper energies (i.e. more and more jammed states), is however still quite controversial. No 
realistic model where this scenario can be proved mathematically yet exists.

The same issues exist also for spin-glasses. Parisi's mean-field solution in this case reveals an even richer 
and more complex landscape structure, with valleys within valleys within valleys, in a hierarchical (fractal) fashion.
Although this fractal picture is very helpful to account for the memory and rejuvenation effects, the way to 
reconcile mean-field with real spin-glasses where the dynamics again becomes slow because 
of the growth of some cooperative length scale is far from settled. A remarkable effect predicted in spin-glasses
is their extreme fragility to tiny temperature changes, that may induce 
large rearrangements in the equilibrium spin configuration.
Such a fragility was also discussed in other contexts (pinned vortex lines, dislocations, domain walls; force chains
in granular materials). The extent to which this `temperature chaos' effect also exists in regular glasses and
allows one to understand rejuvenation in these materials is an open problem.

\section{Open questions}

Therefore, some of the outstanding questions that remain before we can say we understand why glasses do not flow are 
the following: 

\begin{itemize}

\item How relevant (if at all) are mean-field ideas/models for real glasses/spin-glasses? 
Is cooperativity `non topographic' (non thermodynamical) as in mobility defect/facilitated models or
related to an exponential degeneracy of metastable states, as in mean-field models?
Can one make some (controled) 
theoretical progress on a non mean-field model of glasses, or at least formulate a Ginzburg-like criterion to understand the
parameter region where mean-field models are relevant to describe real glassy materials?
Can a frozen, non crystalline thermodynamical glass phase exist in finite dimensions or is it always killed by `activated processes', absent from MCT? How much of 
Parisi's hierarchical energy landscape survives in finite dimensions?  

\item What is the geometry of elementary dynamical excitations in glasses and spin-glasses (strings? fractal clusters?). 
Is there really, experimentally, a detectable growing dynamical length scale in glassy systems (including 
jamming granular materials, soft glassy materials, spin glasses)? How large can this length actually 
grow (probably not more than 10-20), and can this explain the apparent universality of glassy dynamics? Is this length scale important to understand, e.g. 
anomalous phonon modes or fracture in these materials? 

\item Is {\it quenched} disorder crucial or is there a random matrix like theory of the statistics of energy landscapes,
that would make MCT predictions generic? Is the idea of fragility and disorder/temperature chaos, now well established 
for disordered systems (spin glasses, randomly pinned objects) and possibly related to rejuvenation effects, also 
relevant for structural glasses?

\end{itemize}

From a wider perspective, one can ask how much glassy systems, with their profusion of quasi-equilibrium states 
and complex dynamics, can be used as metaphors in other contexts. Combinatorial optimization is an already well 
studied one. Applications in economics/finance/game theory, where equilibrium is often assumed but may never be
reached, is a fascinating prospect.

I want to warmly thank Giulio Biroli for sharing with me his insights on these 
problems, and Piers Coleman for giving me the opportunity to talk at `Frontiers in Condensed Matter 2004' and write this piece. 

\section{(Very) few references}

\begin{itemize} 

\item {\it Some classic papers}: G. Adam, J. H. Gibbs, J. Chem. Phys. {\bf 43} 139 (1965).
M. Goldstein, J. Chem. Phys. {\bf 51}, 3728 (1969).

\item {\it The mean-field scenario for glasses}: T. R. Kirkpatrick, P. G.  Wolynes, Phys. Rev. B {\bf
 36}, 8552 (1987);  T. R. Kirkpatrick, D. Thirumalai, P. G. Wolynes, {\it  Phys. Rev. A} {\bf 40} (1989) 1045; 
 M. M{\'e}zard, {\it First  steps in glass theory}, in ``More is different'', Ong and Bhatt Editors, Princeton University Press (2002).

\item {\it The Mode-Coupling Theory}: W. G{\"o}tze, L. Sj{\"o}gren, Rep. Prog. Phys. {\bf  55} 241 (1992)

\item {\it Length scales within mean-field models}: G. Biroli, J. P. Bouchaud,  {\it Diverging length scale 
and upper critical dimension in the Mode-Coupling Theory of the glass transition}, Europhys. Lett. {\bf 67} 21 (2004);
J. P. Bouchaud, G. Biroli, {\it On the Adam-Gibbs-Kirkpatrick-Thirumalai-Wolynes scenario for the viscosity increase
of glasses}, cond-mat/0406317, to appear in J. Chem. Phys. 
	
\item {\it Non topographic mobility defect models}: J.P. Garrahan, D. Chandler,  Proc. Natl. Acad. Sci. {\bf 100}, 9710 (2003); 
L. Berthier, J.P. Garrahan, Phys. Rev. {\bf E 68}, 041201 (2003)
	
\item {\it Strings and landscape}: M. Vogel, B. Doliwa, A. Heuer and S.C. Glotzer, {\it 
Particle rearrangements during transitions between local minima of the
potential energy landscape}, e-print cond-mat/0309153, 
and refs. therein.
	
\item {\it Length scales and rejuvenation in spin-glasses}: D. S. Fisher, 
D. A. Huse, Phys. Rev B {\bf 38}, 373 (1988).
J.P. Bouchaud, V. Dupuis, J. Hammann, E. Vincent, Phys. Rev B {\bf 65} 
024439 (2001); P. Jonsson, R. Mathieu, P. Nordblad, H. Yoshino, H. Aruga Katori, A. Ito, {\it Spin-glasses, a ghost 
story}, e-print cond-mat/0307640.

\end{itemize}

	\end{document}